\documentclass[12pt]{article}
\input{epsf}

\def\be{\begin{equation}}
\def\ee{\end{equation}}
\def\bea{\begin{eqnarray}}
\def\eea{\end{eqnarray}}
\def\half{\frac{1}{2}}

\begin{document}

\title{\bf Cross-Newell equations for hexagons and triangles}

\author{{\bf Rebecca B. Hoyle} \\
{\it Department of Applied Mathematics and Theoretical Physics,} \\
{\it Silver Street, 
Cambridge CB3 9EW, UK.}}

\maketitle
\begin{abstract}
The Cross-Newell equations for hexagons and triangles are derived for
general real gradient systems, and are found to be in flux-divergence 
form. Specific examples of complex governing
equations that give rise to hexagons and triangles and which have Lyapunov
functionals are also considered, and explicit forms of the Cross-Newell
equations are found in these cases. The general nongradient case is also
discussed; in contrast with the gradient case, the equations are not
flux-divergent. In all cases, the phase stability boundaries and modes of 
instability for general distorted hexagons and triangles can be recovered
from the Cross-Newell equations.
\end{abstract}

\section{Introduction}

Hexagons are a very common planform arising in pattern-forming systems.
The asymmetry between the centres and the edges of the hexagons
leads to a favouring of hexagonal patterns in situations where there is 
intrinsic asymmetry, such as in B\'enard-Marangoni convection~\cite{Ben1}
~\cite{Ben2}~\cite{Ben3} where the 
top surface of the convecting layer is free and the bottom surface is
in contact with a rigid boundary. Most natural systems will have some degree
of asymmetry, and hence hexagons are widely observed, not only in convection 
experiments, but also for example in 
vibrated granular layers~\cite{Melo}~\cite{Umb}
 and during directional solidification~\cite{Morr}. 
Triangular patterns are more unusual, but are seen in some systems, such as
vibrated granular layers~\cite{Umb}.

Cross and Newell~\cite{crnew} pioneered
 a method of describing the behaviour of 
a fully nonlinear roll pattern in an extended system 
by following the evolution of the local phase,
and hence the wavevector, associated with the roll pattern as it varies in
space and time. This method was further developed by Passot and 
Newell~\cite{passnew}
who regularised the Cross-Newell equations outside the region
of roll stability, introducing an order parameter equation to account 
correctly for the behaviour of the pattern in regions where the amplitude is
small. 

The purpose of the current paper is to apply ideas similar to those
of Cross \& Newell~\cite{crnew} and Passot \& Newell~\cite{passnew}
 to the evolution of fully nonlinear
hexagonal and triangular patterns in large aspect ratio systems, such as
those seen in experiments on Rayleigh-B\'enard convection in $\rm SF_6$ near 
the thermodynamical critical point~\cite{Ass}. 

The paper is structured as follows: \S 2 presents a method of deriving the 
Cross-Newell equations for triangles
and hexagons in a general real gradient system. The Cross-Newell
equations for particular complex gradient systems are derived in \S 3 for 
hexagons and \S 4 for triangles. The case of free hexagons and triangles is
discussed in \S 5, and the general nongradient case
in \S 6. Section 7 concludes and indicates some directions for future
investigation.

\section{Derivation of the Cross-Newell equations}

It is assumed that fully developed hexagons or triangles can be described
by a stationary solution $w = w_0({\bf x})$ of an equation $w_t = {\cal L}w
+ {\cal N} w$, where ${\cal L}$ and ${\cal N}$ are linear and nonlinear 
operators respectively, at least one of which is differential, 
with variational 
structure such that 
\be \label{gov}
\int dx dy w_t = - \int dx dy {\delta G \over \delta w},
\ee
where $w$ and $G(w)$ are real.

Hexagons and triangles 
are described by three wavevectors ${\bf k_1}$, ${\bf k_2}$ and ${\bf k_3}$
forming a resonant triad such that ${\bf k_1} + {\bf k_2}+ {\bf k_3}={\bf 0}$.
 In 
the case where the governing equations force this resonance to be maintained, 
the pattern can be described using two phases $\theta_1$ and $\theta_2$ 
associated with 
two of these wavevectors ${\bf k_1} = \nabla \theta_1$ and ${\bf k_2} = 
\nabla \theta_2$. For fully nonlinear triangles and hexagons, the hexagon 
amplitude and the total hexagon phase $a=\theta_1+\theta_2+\theta_3$ (where
${\bf k_3} = -{\bf k_1} -{\bf k_2} = \nabla \theta_3$) are determined 
adiabatically from the two phases $\theta_1$ and $\theta_2$ 
except in the
vicinity of defects where the amplitude is small or when the driving stress
parameter of the system is close
to the critical value for pattern formation so that the amplitude is small
everywhere.
In the `free' case discussed in \S 5, where the resonant triad
may be broken, the hexagon amplitude is slaved to the three independent phases
$\theta_1$, $\theta_2$ and $\theta_3$, except when the amplitude is small.

In a large aspect ratio system, the size and orientation of the hexagons 
will typically change slowly in space and time. To describe these changes, it
is convenient to
introduce large scale phases $\Theta_1 = \epsilon \theta_1$, $\Theta_2 =
\epsilon \theta_2$, where $\epsilon \ll 1$ is the inverse aspect ratio of the 
box, and slow space and time scales ${\bf X} = \epsilon {\bf x}$, $T = 
\epsilon^2 t$. The local wavevectors are then given by
${\bf k_i} = \nabla_{\bf x} \theta_i = \nabla_{\bf X} \Theta_i$, $i=1,2$.

The hexagon solution is now considered to be a function of
the two phases $\theta_1$ and $\theta_2$ and the slow space and time scales,
such that $w\equiv w(\theta_1, \theta_2; {\bf X}, T)$. Hence
the space and time derivatives of $w$ are given by
\bea
\nabla_{\bf x} w(\theta_1, \theta_2; {\bf X}, T) &=&
 ({\bf k_1} \partial_{\theta 1}
+ {\bf k_2} \partial_{\theta 2} + \epsilon \nabla_{\bf X})
w(\theta_1, \theta_2; {\bf X}, T), \\
\partial_t w(\theta_1, \theta_2; {\bf X}, T) &=& (\epsilon \Theta_{1T} 
\partial_{\theta 1} + \epsilon \Theta_{2T} \partial_{\theta 2} + \epsilon^2 
\partial_T) w(\theta_1, \theta_2; {\bf X}, T).
\eea
To leading order then the following equations hold
\bea
w_t &=& \epsilon \{ \Theta_{1T} (\partial_{\theta 1} w_0) + 
\Theta_{2T} (\partial_{\theta 2} w_0) \}, \\
\delta w &=& (\partial_{\theta 1} w_0) \delta \theta_1 + 
(\partial_{\theta 2} w_0) \delta \theta_2.
\eea
Substituting all this information into the governing equation (~\ref{gov})
and averaging over $\theta_1$ and $\theta_2$ gives, to leading order in 
$\epsilon$,
\bea 
&& \epsilon \int dx dy \overline{ \{ \Theta_{1T} (\partial_{\theta 1} w_0) + 
\Theta_{2T} (\partial_{\theta 2} w_0)\} \{ (\partial_{\theta 1} w_0)
 \delta \theta_1
 + (\partial_{\theta 2} w_0) \delta \theta_2 \} }  \nonumber \\
=&&  -\int dx dy {\partial {\overline G} \over \partial k_1^2} \delta k_1^2 
+  {\partial {\overline G} \over \partial k_2^2} \delta k_2^2 
+  {\partial {\overline G} \over \partial {\bf (k_1 . k_2)}} \delta 
({\bf k_1 . k_2}),
\eea
where $\overline{( . )}$ denotes the averaging.
Remarking that $\delta k_1^2 = \delta (\nabla_{\bf x} \theta_1 . 
\nabla_{\bf x} \theta_1) = 
2 \nabla_{\bf x} \theta_1 . \nabla_{\bf x} \delta \theta_1 = 2 {\bf k_1} . 
\nabla_{\bf x} 
\delta \theta_1$,
and similarly that 
$\delta k_2^2 = 2 {\bf k_2} . \nabla_{\bf x} \delta \theta_2$ and
$\delta ({\bf k_1 . k_2}) = {\bf k_1}. \nabla_{\bf x} \delta \theta_2 + 
{\bf k_2}. \nabla_{\bf x} \delta \theta_1$,
the divergence theorem can be used with suitable boundary conditions, to 
show that 
\bea 
&& \epsilon \int dx dy \overline{ \{ \Theta_{1T} (\partial_{\theta 1} w_0) + 
\Theta_{2T} (\partial_{\theta 2} w_0)\} \{ (\partial_{\theta 1} w_0)
 \delta \theta_1
+ (\partial_{\theta 2} w_0) \delta \theta_2 \} }  \nonumber \\
=&&  \epsilon \int dx dy \nabla . \left \{ 2{\bf k_1} 
{\partial {\overline G} \over \partial k_1^2} \right \} \delta \theta_1
+ \nabla . \left \{ 2{\bf k_2}  {\partial {\overline G} \over \partial k_2^2}
\right \} \delta \theta_2 \nonumber \\
&&+ \epsilon \int dx dy  \nabla . \left \{ ({\bf k_1} \delta \theta_2 +
{\bf k_2} \delta \theta_1) 
 {\partial {\overline G} \over \partial {\bf (k_1 . k_2)}} \right \},
\eea
where $\nabla \equiv \nabla_{\bf X} = \epsilon^{-1}(\nabla_{\bf x} - {\bf k_1}
\partial_{\theta 1} - {\bf k_2} \partial_{\theta 2})$.
Since $\delta \theta_1$ and $\delta \theta_2$ are arbitrary, it is possible
to extract
the phase equations
\bea
\Theta_{1T} \overline{|\partial_{\theta 1} w_0|^2} + \Theta_{2T} 
\overline{(\partial_{\theta 1} w_0) (\partial_{\theta 2} w_0)} &=& 
\nabla . \left (2{\bf k_1} {\partial {\overline G} \over \partial k_1^2} +
{\bf k_2}{\partial {\overline G} \over \partial {\bf (k_1 . k_2)}} \right ), 
\label{cn1}\\
\Theta_{2T} \overline{|\partial_{\theta 2} w_0|^2} + \Theta_{1T} 
\overline{(\partial_{\theta 1} w_0) (\partial_{\theta 2} w_0)} &=& 
\nabla . \left (2{\bf k_2} {\partial {\overline G} \over \partial k_2^2} +
{\bf k_1}{\partial {\overline G} \over \partial {\bf (k_1 . k_2)}} \right ).
\label{cn2}
\eea

The phase stability boundaries for general distorted hexagons and triangles
defined by wavevectors ${\bf k_1}$ and ${\bf k_2}$ can be recovered from the 
Cross-Newell equations by first writing the equations explicitly in terms of 
the phases to give 
\bea
&&\Theta_{1T} \overline{|\partial_{\theta 1} w_0|^2} + \Theta_{2T} 
\overline{(\partial_{\theta 1} w_0) (\partial_{\theta 2} w_0)} = \nonumber \\
&&\nabla . \left \{ 2 \nabla \Theta_1 {\partial {\overline G} \over 
\partial k_1^2}(|\nabla \Theta_1|^2, |\nabla \Theta_2|^2, (\nabla \Theta_1.
\nabla \Theta_2)) \right \} + \nonumber \\
&&\nabla . \left \{ 
\nabla \Theta_2 {\partial {\overline G} \over \partial {\bf (k_1 . k_2)}}
(|\nabla \Theta_1|^2, |\nabla \Theta_2|^2, (\nabla \Theta_1.
\nabla \Theta_2)) 
\right \}, \\
&&\Theta_{2T} \overline{|\partial_{\theta 2} w_0|^2} + \Theta_{1T} 
\overline{(\partial_{\theta 1} w_0) (\partial_{\theta 2} w_0)} = \nonumber \\
&&\nabla . \left \{ 2 \nabla \Theta_2 
{\partial {\overline G} \over \partial k_2^2} 
(|\nabla \Theta_1|^2, |\nabla \Theta_2|^2, (\nabla \Theta_1.
\nabla \Theta_2)) \right \} + \nonumber  \\
&&  \nabla . \left \{ 
\nabla \Theta_1 {\partial {\overline G} \over \partial {\bf (k_1 . k_2)}} 
(|\nabla \Theta_1|^2, |\nabla \Theta_2|^2, (\nabla \Theta_1.
\nabla \Theta_2))\right \},
\eea
and then
setting $\Theta_1 = {\bf k_1. X} + \tilde \Theta_1$ and
$\Theta_2 = {\bf k_2. X} + \tilde \Theta_2$, where 
$\tilde \Theta_1$ and $\tilde \Theta_2$ are small. Linearising in 
$\tilde \Theta_1$ and $\tilde \Theta_2$, and setting $\tilde \Theta_1 =
\hat \Theta_1 e^{\sigma T + i{\bf \tilde k . X}}$ and $\tilde \Theta_2 = 
\hat \Theta_2 e^{\sigma T + i{\bf \tilde k . X}}$, with $\hat \Theta_1$ 
and $\hat 
\Theta_2$ real constants gives a dispersion relation for the 
growth-rate eigenvalues $\sigma$. Hence the stability boundaries and
modes of instability can be found as in~\cite{me1}~\cite{me2}. Direct 
numerical integration of the Cross-Newell equations could also be used to
determine the region of stable hexagons and triangles, and comparison
could be made with the stability region for regular hexagons found by
other numerical methods as in~\cite{beste}.  

Ideally the governing equations should be real, as assumed here, in order to 
allow the formation of disclinations on an individual set of 
rolls~\cite{passnew}. However,
there do not appear to be simple examples of real governing
equations which give fully nonlinear hexagons or triangles 
as an exact stationary solution, and so
in order to make further explicit analytical 
progress we shift our attention in the 
following section to complex governing equations which do indeed give hexagons.
This is perhaps less of a handicap than it would be in the case of rolls, 
since the canonical hepta-penta defect of hexagons is made up of dislocations,
which can be described by a complex order parameter.  

\section{Cross-Newell equations for hexagons}

With slight modifications to the spatial derivative terms,
 the standard complex amplitude equations for 
hexagons~\cite{newhite}~\cite{gob}
can be used as the basic governing equations, giving
\be
{\partial z_i \over \partial t} = \lambda z_i + \alpha z_{i+1}^*  
z_{i+2}^* - \beta |z_i|^2 z_i -\gamma (|z_{i+1}|^2 +|z_{i+2}|^2)z_i + 
\nabla^2 z_i,
\ee
where $\lambda$, $\alpha$, $\beta$ and $\gamma$ 
are real constants, and where $*$ denotes complex conjugation.  
 The hexagon solutions are represented by $w = Re(z_1+z_2+z_3)$,
$z_i = R_i e^{i\theta_i}$, with $i=1,2,3$ and cyclic. 
Here the usual spatial derivatives have been replaced
by $\nabla^2$ in order to preserve the isotropy of the system.

There is a Lyapunov functional associated with the amplitude equations,
given by
\bea
L &=& - \lambda (|z_1|^2+|z_2|^2+|z_3|^2) - \alpha (z_1z_2z_3+z_1^*z_2^*z_3^*)
+\half \beta (|z_1|^4+|z_2|^4+|z_3|^4) \nonumber \\
&&+ \gamma (|z_1|^2|z_2|^2 + 
|z_2|^2|z_3|^2+|z_3|^2|z_1|^2) + |\nabla z_1|^2 + |\nabla z_2|^2 + 
|\nabla z_3|^2,
\eea
such that
\be
{\partial z_i \over \partial t} = -{\delta L \over \delta z_i^*}.
\ee
There are wavevectors associated with the phases according to ${\bf k_i} = 
\nabla \theta_i$ as before. A hexagonal or triangular pattern arises when 
the sum of the three wavevectors is zero, i.e. $\bf k_1+k_2+k_3=0$.
Hence the total phase $\sum_i \theta_i \equiv a(t)$ is a function
of time only.

The fully nonlinear hexagonal solution takes the form $w=R_1 \cos \theta_1
+ R_2 \cos \theta_2 + R_3 \cos \theta_3$ where
\bea 
&&0 = R_1(\lambda-k_1^2) +\alpha R_2R_3 \cos a - \beta R_1^3 - \gamma (R_2^2+
R_3^2)R_1,  \label{amps}\\
&&0 = R_2(\lambda-k_2^2) +\alpha R_3R_1 \cos a - \beta R_2^3 - \gamma (R_3^2+
R_1^2)R_2, \\
&&0 = R_3(\lambda-k_3^2) +\alpha R_1R_2 \cos a - \beta R_3^3 - \gamma (R_1^2+
R_2^2)R_3, \label{ampsp}\\
&&0 = \alpha \sin a \label{ampsf}
\eea
hold, and where the $R_i$ are nonzero constants. 
Clearly if $\alpha$ is nonzero, as assumed in this section, the
total phase $a$ must take the value $0$ or $\pi$. 
If $\alpha$ is zero, $a$ can take any value.

In the case of nonzero $\alpha$, there are only two independent phases, 
which without
loss of generality are taken to be $\theta_1$ and
$\theta_2$. The third phase $\theta_3=a-\theta_1-\theta_2$ is then dependent,
since $a$ is fixed.

As in the previous section, it is assumed that the wavevectors vary 
slowly in space and time, so that it is possible to define large scale phases
$\Theta_i = \epsilon \theta_i$ and long space and time scales
 such that $\nabla_{\bf x} = {\bf k_1} \partial_{\theta_1}
+ {\bf k_2} \partial_{\theta_2} + \epsilon \nabla_{\bf X}$ and
 $\partial_t = \epsilon ( \Theta_{1T} \partial_{\theta_1}+\Theta_{2T}
 \partial_{\theta_2})$. The solution is expanded in the form
${\bf z}\equiv (z_1,z_2,z_3) = {\bf z_0} + \epsilon {\bf \tilde z_1} +
\epsilon^2 {\bf \tilde z_2} +...$, where ${\bf z_0}$ is the fully nonlinear
hexagon solution above.

To leading order, the Lyapunov functional for the fully nonlinear hexagons 
takes the form
\bea \label{lap}
L &=& -\lambda (R_1^2+R_2^2+R_3^2) - 2 \alpha R_1R_2R_3 \cos a + \half 
(R_1^4+R_2^4+R_3^4) \nonumber \\
&&+ \gamma(R_1^2R_2^2+R_2^2R_3^2+R_3^2R_1^2) + k_1^2 R_1^2
+k_2^2 R_2^2 + k_3^2 R_3^2.
\eea
Since $\bf k_1+k_2+k_3=0$ holds, $k_3^2$ can be rewritten $k_1^2+k_2^2+2{\bf 
k_1 . k_2}$, and it is clear that the $R_i$, $a$ and $L$ all depend only on 
$k_1^2$, $k_2^2$ and ${\bf k_1 . k_2}$. Hence the variation $\delta L$ in the
Lyapunov functional is given by
\be
\delta L = {\partial L \over \partial k_1^2} \delta k_1^2 + 
{\partial L \over \partial k_2^2} \delta k_2^2 +
{\partial L \over \partial {\bf (k_1 . k_2)}} \delta {\bf (k_1 . k_2)}.
\ee
It is also clear that 
\be \label{dzs}
\delta L = {\partial L \over \partial z_i^*}\delta z_i^*
+ {\partial L \over \partial z_i}\delta z_i = - {\partial z_i \over \partial
t}\delta z_i^* -  {\partial z_i^* \over \partial
t}\delta z_i 
\ee
holds. 
Further, it can be seen that 
\bea
\delta {\bf z_0^*} =&& 
(\partial_{\theta_1}
{\bf z_0^*}) \delta \theta_1 + (\partial_{\theta_2}
{\bf z_0^*}) \delta \theta_2, \\ 
\partial_t {\bf z} =&& \epsilon \Theta_{1T}
(\partial_{\theta_1} {\bf z_0}) + \epsilon \Theta_{2T}
(\partial_{\theta_2} {\bf z_0}) + O(\epsilon^2)
\eea
 hold. Substituting these
into equation (~\ref{dzs}) gives
\be
\delta L = -\epsilon \{ \Theta_{1T}
(\partial_{\theta_1} {\bf z_0}) + \Theta_{2T}
(\partial_{\theta_2} {\bf z_0})\}.\{( \partial_{\theta_1}
{\bf z_0^*}) \delta \theta_1 + (\partial_{\theta_2}
{\bf z_0^*}) \delta \theta_2 \} + c.c.
\ee
to leading order, where $c.c.$ denotes complex conjugate. Considering $\int 
\delta L dx dy$, where the integral is taken over the 
whole domain, gives
\bea \label{dls}
&&\int dx dy \epsilon \{ \Theta_{1T}
(\partial_{\theta_1} {\bf z_0}) + \Theta_{2T}
(\partial_{\theta_2} {\bf z_0}) \}. \{ ( \partial_{\theta_1}
{\bf z_0^*}) \delta \theta_1 + (\partial_{\theta_2}
{\bf z_0^*}) \delta \theta_2 \} + c.c. \nonumber \\
&&= -\int dx dy {\partial L \over \partial k_1^2} \delta k_1^2 + 
{\partial L \over \partial k_2^2} \delta k_2^2 +
{\partial L \over \partial {\bf (k_1 . k_2)}} \delta {\bf (k_1 . k_2)}
\nonumber \\
&&=\epsilon \int dx dy \nabla . \left ( 2{\bf k_1}{\partial L \over \partial 
k_1^2} + 
{\bf k_2}{\partial L \over \partial {\bf (k_1 . k_2)}}\right ) 
\delta \theta_1 +
\nabla . \left ( 2{\bf k_2}{\partial L \over \partial k_2^2} + 
{\bf k_1}{\partial L \over \partial {\bf (k_1 . k_2)}}\right ) \delta \theta_2,
\nonumber \\
\eea
from which it is easy to identify the phase equations
\bea
\Theta_{1T} (\partial_{\theta_1} {\bf z_0}) . (\partial_{\theta_1}
{\bf z_0^*}) + \Theta_{2T} (\partial_{\theta_2} {\bf z_0}) . 
(\partial_{\theta_1} {\bf z_0^*}) + c.c. &=& 
 \nabla . 
\left ( 2{\bf k_1}{\partial L \over \partial k_1^2} + 
{\bf k_2}{\partial L \over \partial {\bf (k_1 . k_2)}}\right ), \nonumber \\
\\
\Theta_{2T} (\partial_{\theta_2} {\bf z_0}) . (\partial_{\theta_2} 
{\bf z_0^*}) + \Theta_{1T} (\partial_{\theta_1} {\bf z_0}) . 
(\partial_{\theta_2} {\bf z_0^*}) + c.c. &=&
\nabla . 
\left ( 2{\bf k_2}{\partial L \over \partial k_2^2} + 
{\bf k_1}{\partial L \over \partial {\bf (k_1 . k_2)}}\right ). \nonumber \\
\eea
The straightforward substitutions 
\bea
{\bf z_0} = \left ( \matrix{R_1 e^{i\theta_1} \cr R_2 e^{i\theta_2} \cr
R_3 e^{i(a-\theta_1-\theta_2)}} \right ), \\
\partial_{\theta_1} {\bf z_0} = \left ( \matrix{iR_1 e^{i\theta_1} \cr 0 \cr
-iR_3 e^{i(a-\theta_1-\theta_2)}} \right ), \\
\partial_{\theta_2} {\bf z_0} = \left ( \matrix{0\cr iR_2 e^{i\theta_2} \cr
-iR_3 e^{i(a-\theta_1-\theta_2)}} \right ), 
\eea
reduce the phase equations to
\bea
2(R_1^2+R_3^2)\Theta_{1T} + 2R_3^2 \Theta_{2T} = 
\nabla . \left ( 2 {\bf k_1} {\partial L \over \partial k_1^2} 
+ {\bf k_2}{\partial L \over \partial {\bf (k_1 . k_2)}}\right ), \\
2(R_2^2+R_3^2)\Theta_{2T} + 2R_3^2 \Theta_{1T} = 
\nabla . \left ( 2 {\bf k_2} {\partial L \over \partial k_2^2} 
+ {\bf k_1}{\partial L \over \partial {\bf (k_1 . k_2)}}\right ).
\eea
Substituting equations (~\ref{amps})-(~\ref{ampsf}) into equation (~\ref{lap})
 gives a 
simpler expression for the Lyapunov functional
\be
L = \alpha R_1 R_2R_3 \cos a - \gamma (R_1^2R_2^2 + R_2^2R_3^2+R_3^2 R_1^2)
- \half \beta (R_1^4+R_2^4+R_3^4),
\ee
which when differentiated with respect to $k_1^2$ becomes
\bea
{\partial L \over \partial k_1^2} &=& \alpha \cos a \left
( {\partial R_1 \over \partial  k_1^2} R_2R_3 +  R_1 {\partial R_2 \over 
\partial  k_1^2} R_3 + R_1 R_2{\partial R_3 \over \partial  k_1^2} \right )
\nonumber \\
&&-2\gamma\left (R_1 {\partial R_1 \over \partial  k_1^2} 
 (R_2^2+R_3^2)
+ R_2 {\partial R_2 \over \partial  k_2^2}(R_3^2+R_1^2)  
+ R_3 {\partial R_3 \over \partial  k_2^2}(R_1^2+R_2^2) \right ) \nonumber \\
&&-2\beta \left ( R_1^3 {\partial R_1 \over \partial  k_1^2} +  R_2^3
 {\partial R_2 \over \partial  k_1^2} + R_3^3 {\partial R_3 \over \partial  
k_1^2} \right ).
\eea
Dividing the amplitude equations (~\ref{amps})-(~\ref{ampsp})
 through by $R_1$, $R_2$, $R_3$
respectively and then differentiating gives
\bea
0&=&-1 + {\alpha \cos a \over  R_1^2} \left (R_1R_2 {\partial R_3  
\over \partial k_1^2}
+ R_3R_1 {\partial R_2 \over \partial k_1^2} - R_2R_3 {\partial R_1 \over 
\partial k_1^2}
\right ) - 2\beta R_1 {\partial R_1 \over \partial k_1^2} \nonumber \\
&&- 2\gamma \left (
R_2 {\partial R_2 \over \partial k_1^2} + R_3 {\partial R_3 \over \partial
k_1^2} \right ), \\
0&=&{\alpha \cos a \over R_2^2} \left (R_2R_3 {\partial R_1 \over \partial
k_1^2}
+ R_1R_2 {\partial R_3 \over \partial k_1^2} - R_3R_1 {\partial R_2 \over 
\partial k_1^2}
\right ) - 2\beta R_2 {\partial R_2 \over \partial k_1^2} \nonumber \\
&&- 2\gamma \left (
R_3 {\partial R_3 \over \partial k_1^2} + R_1 {\partial R_1 \over \partial
k_1^2} \right ), \\
0&=&-1 + {\alpha \cos a \over R_3^2} \left (R_3R_1 {\partial R_2 \over 
\partial k_1^2}
+ R_2R_3 {\partial R_1 \over \partial k_1^2} - R_1R_2 {\partial R_3 \over 
\partial k_1^2}
\right ) - 2\beta R_3 {\partial R_3 \over \partial k_1^2} \nonumber \\
&&- 2\gamma \left (
R_1 {\partial R_1 \over \partial k_1^2} + R_2 {\partial R_2 \over \partial 
k_1^2} \right ), 
\eea
which when multiplied by $R_1^2$, $R_2^2$, $R_3^2$ respectively and added 
give
\bea
0&=&-(R_1^2+R_3^2) +\alpha \cos a \left
( {\partial R_1 \over \partial  k_1^2} R_2R_3 +  R_1 {\partial R_2 \over 
\partial  k_1^2} R_3 + R_1 R_2{\partial R_3 \over \partial  k_1^2} \right )
\nonumber \\
&& -2\gamma\left (R_1 {\partial R_1 \over \partial  k_1^2} 
 (R_2^2+R_3^2)
+ R_2 {\partial R_2 \over \partial  k_2^2}(R_3^2+R_1^2) + 
+ R_3 {\partial R_3 \over \partial  k_2^2}(R_1^2+R_2^2) \right ) \nonumber \\
&&-2\beta \left ( R_1^3 {\partial R_1 \over \partial  k_1^2} +  R_2^3
 {\partial R_2 \over \partial  k_1^2} + R_3^3 {\partial R_3 \over \partial  
k_1^2} \right ),
\eea
and hence
\be 
{\partial L \over \partial k_1^2} = (R_1^2+R_3^2).
\ee
Similarly it is found that 
\bea
{\partial L \over \partial k_2^2} &=& (R_2^2+R_3^2), \\
{\partial L \over \partial {\bf (k_1 . k_2)}} &=& 2R_3^2,
\eea
and hence
\bea
(R_1^2+R_3^2)\Theta_{1T} + R_3^2 \Theta_{2T} =
\nabla . \left ( {\bf k_1} (R_1^2+R_3^2) + {\bf k_2} R_3^2 \right ), \\
(R_2^2+R_3^2)\Theta_{2T} + R_3^2 \Theta_{1T} = 
\nabla . \left ( {\bf k_2} (R_2^2+R_3^2) + {\bf k_1} R_3^2 \right ),
\eea
which gives the Cross-Newell equations
\bea
(R_1^2R_2^2+R_2^2R_3^2+R_3^2R_1^2)\Theta_{1T} &=& R_2^2 \nabla .
({\bf k_1} R_1^2 - {\bf k_3} R_3^2) + R_3^2 \nabla .
({\bf k_1} R_1^2 - {\bf k_2} R_2^2), \nonumber \\
\\
(R_1^2R_2^2+R_2^2R_3^2+R_3^2R_1^2)\Theta_{2T} &=& R_1^2 \nabla .
({\bf k_2} R_2^2 - {\bf k_3} R_3^2) + R_3^2 \nabla .
({\bf k_2} R_2^2 - {\bf k_1} R_1^2) \nonumber \\
\eea
upon rearrangement.

\section{Cross-Newell equations for triangles}

A similar approach can be adopted to derive Cross-Newell equations 
for triangles starting from the governing equations
\bea
{\partial z_i \over \partial t} &=& z_i \{\lambda - a_1 |z_i|^2 
-a_2 ( |z_1|^2 + |z_2|^2 + |z_3|^2 ) -a_3 (z_1z_2z_3+z_1^* z_2^*
z_3^*)\} \nonumber \\
&&+z_{i+1}^* z_{i+2}^*\{ \delta -a_3(|z_1|^2+|z_2|^2
+|z_3|^2)\}+ \nabla^2 z_i.
\eea
These are the lowest order amplitude equations that permit triangles as a
stationary solution~\cite{gob} and once again the spatial derivatives have
been chosen to ensure that the governing equations are isotropic.
Fully nonlinear stationary triangles satisfy
\bea
0 &=& \lambda -k_i^2 -a_1|z_i|^2 -a_2(|z_1|^2 + |z_2|^2 +|z_3|^2)
-a_3(z_1z_2z_3 + z_1^* z_2^* z_3^*), \\
0 &=& \delta - a_3(|z_1|^2 + |z_2|^2 +|z_3|^2).
\eea
Writing $z_i = R_i e^{i \theta_i}$ as before with $a= \theta_1+\theta_2
+ \theta_3$, gives 
\bea
R_1^2 &=& {(-2k_1^2 + k_2^2 + k_3^2) \over 3 a_1} + {\delta \over 3 a_3}, \\
R_2^2 &=& {(k_1^2 - 2k_2^2 + k_3^2) \over 3 a_1} + {\delta \over 3 a_3}, \\
R_3^2 &=& {(k_1^2 + k_2^2 - 2k_3^2) \over 3 a_1} + {\delta \over 3 a_3}, \\
\cos a &=& { 1 \over 6a_3 R_1R_2R_3} \left ( 3\lambda - (k_1^2 + k_2^2 + k_3^2)
- {(a_1 + 3a_2)\delta \over a_3} \right ).
\eea

The Lyapunov functional is given by
\bea
L &=& -\lambda (|z_1|^2 + |z_2|^2 + |z_3|^2) + \half (a_1+a_2)
(|z_1|^4 + |z_2|^4 + |z_3|^4) \nonumber \\
&&+ a_2(|z_1|^2|z_2|^2 + |z_2|^2|z_3|^2
+|z_3|^2|z_1|^2) \\
&&+ a_3(|z_1|^2 + |z_2|^2 + |z_3|^2)(z_1z_2z_3 + 
z_1^*z_2^*z_3^*) \nonumber \\
&&- \delta (z_1z_2z_3 + z_1^*z_2^*z_3^*)
+ |\nabla z_1|^2 +  |\nabla z_2|^2 + |\nabla z_3|^2,
\eea
which gives, upon substitution for $R_i$ and $a$, 
\bea
L &=& -{ 1 \over 3 a_1} \{k_1^4 +k_2^4 + 4({\bf k_1} . {\bf k_2})^2 +k_1^2k_2^2
-2k_1^2({\bf k_1} . {\bf k_2})-2k_2^2({\bf k_1} . {\bf k_2})\} \nonumber \\
 &&+ { 2\delta \over a_3}(k_1^2+k_2^2 + {\bf k_1} . {\bf k_2})
- {3 \delta \lambda \over a_3} + {5 \delta^2 
(a_1 + 3a_2) \over 6 a_3^2}.
\eea
As before the phase equations are given by
\bea
2(R_1^2+R_3^2)\Theta_{1T} + 2R_3^2 \Theta_{2T} = 
\nabla . \left ( 2 {\bf k_1} {\partial L \over \partial k_1^2} 
+ {\bf k_2}{\partial L \over \partial {\bf (k_1 . k_2)}}\right ), \\
2(R_2^2+R_3^2)\Theta_{2T} + 2R_3^2 \Theta_{1T} = 
\nabla . \left ( 2 {\bf k_2} {\partial L \over \partial k_2^2} 
+ {\bf k_1}{\partial L \over \partial {\bf (k_1 . k_2)}}\right ).
\eea
It is easily seen that 
\bea
{\partial L \over \partial k_1^2} &=& -{1 \over 3 a_1}(2k_1^2 - k_2^2 + 
2 {\bf k_1 . k_2}) + {2\delta \over a_3}, \\
{\partial L \over \partial k_2^2} &=& -{1 \over 3 a_1}(-k_1^2 + 2k_2^2 + 
2 {\bf k_1 . k_2}) + {2\delta \over a_3}, \\
{\partial L \over \partial ({\bf k_1 . k_2})} &=& -{2 \over 3 a_1}
(4{\bf k_1 . k_2} + k_1^2 + k_2^2) +{2\delta \over a_3}.
\eea
After some rearrangements and substitutions the Cross-Newell equations for 
triangles are found to be
\bea
&&(R_1^2R_2^2+R_2^2R_3^2+R_3^2R_1^2)\Theta_{1T} = \nonumber \\
&&R_2^2 \nabla . ({\bf k_1}
(R_1^2 + 2\delta / 3 a_3)-{\bf k_3}(R_3^2 + 2\delta/3a_3)) \nonumber \\
&&+R_3^2 \nabla . ({\bf k_1}
(R_1^2 + 2\delta / 3 a_3)-{\bf k_2}(R_2^2 + 2\delta/3a_3)), \\
&&(R_1^2R_2^2+R_2^2R_3^2+R_3^2R_1^2)\Theta_{2T} = \nonumber \\
&& R_3^2 \nabla . ( {\bf k_2}
(R_2^2 + 2\delta / 3 a_3)-{\bf k_1}(R_1^2 + 2\delta/3a_3)) \nonumber \\
&&+ R_1^2 \nabla . ( {\bf k_2}(R_2^2 + 2\delta / 3 a_3) -
{\bf k_3}(R_3^2+ 2\delta / 3 a_3)) .
\eea

\section{Free hexagons and triangles}

The Cross-Newell equations take a different form when the total phase
$a$ is not constrained to remain fixed by the governing equations,
for example in the case $\alpha=0$ in \S 3.
All three phases $\theta_i$ are independent, with the following consequent 
modifications of the analysis 
\bea
\nabla_{\bf x} &=& {\bf k_1} \partial_{\theta_1}
+ {\bf k_2} \partial_{\theta_2} + {\bf k_3} \partial_{\theta_3} 
+ \epsilon \nabla_{\bf X}, \\
\partial_t &=& \epsilon (\Theta_{1T} \partial_{\theta_1}
+\Theta_{2T} \partial_{\theta_2}+\Theta_{3T}
 \partial_{\theta_3}),\\
\delta {\bf z_0^*} &=& 
\partial_{\theta_1}
{\bf z_0^*} \delta \theta_1 + \partial_{\theta_2}
{\bf z_0^*} \delta \theta_2 + \partial_{\theta_3}
{\bf z_0^*} \delta \theta_3,
\eea
which lead to the phase equations
\bea
\Theta_{1T} \overline{|\partial_{\theta 1} w_0|^2} = \nabla . \left ( 2 
{\bf k_1} {\partial \overline{G} \over \partial k_1^2} \right ), \label{free1}
\\
\Theta_{2T} \overline{|\partial_{\theta 2} w_0|^2} = \nabla . \left ( 2 
{\bf k_2} {\partial \overline{G} \over \partial k_2^2} \right ), \label{free2}
\\
\Theta_{3T} \overline{|\partial_{\theta 3} w_0|^2} = \nabla . \left ( 2 
{\bf k_3} {\partial \overline{G} \over \partial k_3^2} \right ). \label{free3}
\eea
In the hexagon case of \S 3, the Cross-Newell equations would be
\bea
R_1^2 \Theta_{1T} = \nabla . (R_1^2 {\bf k_1}), \label{freec1} \\
R_2^2 \Theta_{2T} = \nabla . (R_2^2 {\bf k_2}), \label{freec2} \\
R_3^2 \Theta_{3T} = \nabla . (R_3^2 {\bf k_3}). \label{freec3}
\eea 
These are the equations that would have been found for three independent
sets of rolls in the same system, as might have been expected, since nothing in
the analysis constrains the hexagons or triangles to remain hexagonal or
triangular. In particular, it is not to be expected that the 
condition ${\bf k_1 + k_2 +
k_3 = 0}$ will be maintained over long times.

For hexagons which remain everywhere exactly hexagonal, 
such that $|{\bf k_1}|=|{\bf k_2}|=|{\bf k_3}|=k$ and $R_1=R_2=R_3\equiv R$, 
the phase equations also take the form (~\ref{freec1})-(~\ref{freec3}), since
the size and orientation of a hexagon can then be determined from a
single wavevector, as in the roll case. However, this constrains the 
hexagons to behave as a rotating, shrinking or expanding 
lattice, which is clearly not a realistic model for most experiments.

\section{Flux-divergence form and the general nonvariational case}

\begin{figure}
\centerline{\epsffile{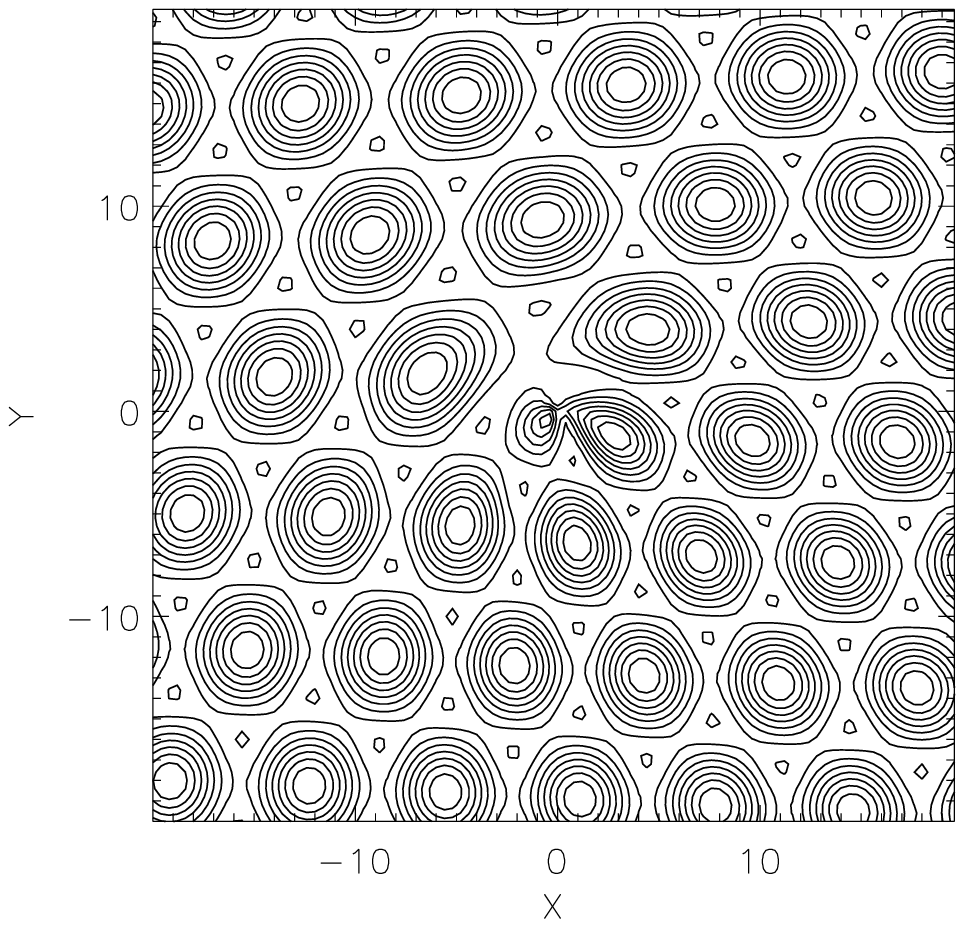}}
\caption{\em A harmonic hepta-penta defect. The figure shows contours of 
$f({\bf X})=\cos \Theta_1+ \cos \Theta_2 + \cos \Theta_3$, where $\Theta_1 =
R \cos (\alpha+\pi/2)-(\alpha+\pi/2)$, $\Theta_2 =
R \cos (\alpha+\pi/6)-(\alpha+\pi/6)$, $\Theta_3 = -\Theta_1 - \Theta_2$,
and where ($R$, $\alpha$) are polar coordinates for ${\bf X}$.}
\label{fig1}
\end{figure}

The Cross-Newell equations (~\ref{cn1}) and (~\ref{cn2}) for hexagons and 
triangles in gradient systems are in flux-divergence form, which has
consequences for defect formation, as in the case of 
rolls~\cite{passnew}.
Note that stationary solutions of equations (~\ref{cn1}) and (~\ref{cn2})
take the form 
\bea
\nabla . \left (2{\bf k_1} {\partial {\overline G} \over \partial k_1^2} +
{\bf k_2}{\partial {\overline G} \over \partial {\bf (k_1 . k_2)}} \right ) 
&&=0, \\
\nabla . \left (2{\bf k_2} {\partial {\overline G} \over \partial k_2^2} +
{\bf k_1}{\partial {\overline G} \over \partial {\bf (k_1 . k_2)}} \right )
&&=0.
\eea
Following~\cite{passnew}, it is interesting to set 
\be
{\partial {\overline G} \over \partial k_1^2}=
{\partial {\overline G} \over \partial k_2^2}=
{\partial {\overline G} \over \partial {\bf (k_1 . k_2)}}=1,
\ee
which implies that ${\bf \nabla .  k_1= \nabla . k_2=0}$ hold. 
Since the wavevectors are gradients of the phases, it is clear that
 $\nabla \times {\bf k_1} = \nabla \times {\bf k_2}=0$ also hold, and 
that $\nabla^2 \Theta_1 =\nabla^2 \Theta_2=0$. The solutions of these
equations are the harmonic defects catalogued in~\cite{passnew}.
Despite being energetically unreasonable, and hence looking somewhat 
unphysical, because they contain features at 
wavelengths which lie outside the stable region,
they provide a good illustration
of the topology of real defects. 
It is possible to construct a harmonic hepta-penta defect by positioning two 
harmonic dislocations~\cite{passnew} on top of each other, as shown in 
figure~\ref{fig1}. The hepta-penta defect is the canonical defect of hexagons.

In the general nonvariational case, the Cross-Newell equations can be 
written
\bea
\tau(k_1^2,k_2^2,{\bf k_1 . k_2}) \Theta_{1T} &=& 
\alpha_1(k_1^2,k_2^2,{\bf k_1 . k_2}) \nabla . {\bf k_1} 
+\alpha_2(k_1^2,k_2^2,{\bf k_1 . k_2}) {\bf k_1}.\nabla k_1 \nonumber \\
&&+\alpha_3(k_1^2,k_2^2,{\bf k_1 . k_2}) {\bf k_1}.\nabla k_2
+\alpha_4(k_1^2,k_2^2,{\bf k_1 . k_2}) {\bf k_1}.\nabla {\bf (k_1. k_2)} 
\nonumber \\
&&+\beta_1(k_1^2,k_2^2,{\bf k_1 . k_2}) \nabla . {\bf k_2}
+\beta_2(k_1^2,k_2^2,{\bf k_1 . k_2}) {\bf k_2}.\nabla k_1 \nonumber \\
&&+\beta_3(k_1^2,k_2^2,{\bf k_1 . k_2}) {\bf k_2}.\nabla k_2
+\beta_4(k_1^2,k_2^2,{\bf k_1 . k_2}) {\bf k_2}.\nabla {\bf (k_1. k_2)}, 
\nonumber \\
&& \\
\tau(k_2^2,k_1^2,{\bf k_1 . k_2}) \Theta_{2T} &=& 
\alpha_1(k_2^2,k_1^2,{\bf k_1 . k_2}) \nabla . {\bf k_2} 
+\alpha_2(k_2^2,k_1^2,{\bf k_1 . k_2}) {\bf k_2}.\nabla k_2 \nonumber \\
&&+\alpha_3(k_2^2,k_1^2,{\bf k_1 . k_2}) {\bf k_2}.\nabla k_1
+\alpha_4(k_2^2,k_1^2,{\bf k_1 . k_2}) {\bf k_2}.\nabla {\bf (k_1. k_2)} 
\nonumber \\
&&+\beta_1(k_2^2,k_1^2,{\bf k_1 . k_2}) \nabla . {\bf k_1} 
+\beta_2(k_2^2,k_1^2,{\bf k_1 . k_2}) {\bf k_1}.\nabla k_2 \nonumber \\
&&+\beta_3(k_2^2,k_1^2,{\bf k_1 . k_2}) {\bf k_1}.\nabla k_1
+\beta_4(k_2^2,k_1^2,{\bf k_1 . k_2}) {\bf k_1}.\nabla {\bf (k_1. k_2)}.
\nonumber \\
\eea
In contrast to the variational case, these equations cannot in general be 
reduced to flux-divergence form, 
and hence 
it cannot be assumed that such general hexagonal patterns 
will have defects whose
topology is given by that of harmonic defects.

\section{Conclusion}

This paper has derived the Cross-Newell equations for triangles and hexagons in
a general real gradient system. The resulting equations can be put into
flux-divergence form, indicating that the topology of defects of such a
hexagonal pattern can be described by that of harmonic defects~\cite{passnew}.
The general
nonvariational case, however, is not flux-divergent. In both cases, the 
phase stability boundaries and modes of instability for general distorted
hexagons and triangles can be recovered from
the Cross-Newell equations.

An explicit analytical form for the Cross-Newell 
equations is found for both hexagons and
triangles in the case where the
governing equations are generalisations of the corresponding complex
amplitude equations. 

This work suggests avenues for further investigation. In particular, 
it would be interesting to analyse the Cross-Newell equations in a 
general nonvariational system, and also to integrate the phase equations 
numerically in order to make a comparison with an integration of the 
full governing equations, for example in order to compare the regions
of stability of hexagons. A further interesting possibility is to
use the Cross-Newell equations to investigate the simultaneous occurrence 
of up- and down-hexagons~\cite{dewel}. These avenues will form the 
basis of future work.

\smallskip

\leftline{\large \bf Acknowledgements}

\noindent The author would like to thank Alan Newell and Thierry Passot for 
interesting and helpful discussions. 

This work was supported by King's College, Cambridge.

\vfil\eject

\end{document}